\documentclass[a4paper,fleqn,usenatbib]{mnras}

\usepackage{newtxtext,newtxmath}

\usepackage[T1]{fontenc}
\usepackage{ae,aecompl}

\usepackage{graphicx}	
\usepackage{amsmath}	
\usepackage{amssymb}	
\usepackage[para,online,flushleft]{threeparttable}
\usepackage{color, soul}

\title[Project 1640: HD 114174 B]{Project 1640 Observations of the White Dwarf HD 114174 B}

\author[E. Bacchus]{
E. Bacchus,$^{1}$\thanks{E-mail: ebacchus@ast.cam.ac.uk}
I. R. Parry,$^{1}$
R. Oppenheimer,$^{2}$
J. Aguilar,$^{3}$
C. Beichman,$^{4,5,6}$
\newauthor
D. Brenner,$^{2}$
R. Burruss,$^{4}$
E. Cady,$^{4}$
S. Luszcz-Cook,$^{2}$
J. Crepp,$^{7}$
R. Dekany,$^{5}$
\newauthor
A. Gianninas,$^{8}$
L. Hillenbrand$^{5}$
M. Kilic,$^{8}$
D. King,$^{1}$
T. G. Lockhart,$^{4}$
C. T. Matthews,$^{7}$
\newauthor
R. Nilsson,$^{2,5,9}$
L. Pueyo,$^{3,10}$
E. L. Rice,$^{2}$
L. C. Roberts, Jr.,$^{4}$
A. Sivaramakrishnan,$^{10}$
\newauthor
R. Soummer,$^{10}$
G. Vasisht,$^{4}$
A. Veicht,$^{2}$
C. Zhai$^{4}$
and N. T. Zimmerman$^{11}$
\\
\\
$^{1}$Institute of Astronomy, University of Cambridge, Madingley Rd, Cambridge, UK, CB3 0HA\\
$^{2}$AMNH, Central Park West at 79th Street, New York, NY 10024-5192\\
$^{3}$Johns Hopkins University, 3400 N. Charles Street, Baltimore, MD 21218, USA\\
$^{4}$Jet Propulsion Laboratory, California Institute of Technology, 4800 Oak Grove Drive, Pasadena CA 91109, USA\\
$^{5}$Division of Physics, Mathematics, and Astronomy, California Institute of Technology, Pasadena, CA 91125, USA\\
$^{6}$NASA Exoplanet Science Institute, 770 S. Wilson Avenue, Pasadena, CA 911225, USA\\
$^{7}$Department of Physics, University of Notre Dame, 225 Nieuwland Science Hall, Notre Dame, IN, 46556, USA\\
$^{8}$Department of Physics and Astronomy, University of Oklahoma, Norman, OK 73019, USA\\
$^{9}$Department of Astronomy, Stockholm University, AlbaNova University Center, Roslagstullsbacken 21, SE-106 91 Stockholm, Sweden\\
$^{10}$Space Telescope Science Institute, 3700 San Martin Drive, Baltimore, MD 21218, USA\\
$^{11}$NASA Goddard Space Flight Center, Exoplanets and Stellar Astrophysics Laboratory, 8800 Greenbelt Rd, Greenbelt MD 20771, USA\\
}
\date{Accepted XXX. Received YYY; in original form ZZZ}

\pubyear{2016}

\begin{document}
\label{firstpage}
\pagerange{\pageref{firstpage}--\pageref{lastpage}}
\maketitle

\begin{abstract}
We present the first near infra-red spectrum of the faint white dwarf companion HD 114174 B, obtained with Project 1640.  Our spectrum, covering the $Y$, $J$ and $H$ bands, combined with previous TRENDS photometry measurements, allows us to place further constraints on this companion.  We suggest two possible scenarios; either this object is an old, low mass, cool H atmosphere white dwarf with $T_\text{eff} \sim 3800$ K or a high mass white dwarf with $T_\text{eff} > 6000$ K, potentially with an associated cool ($T_\text{eff} \sim 700$ K) brown dwarf or debris disk resulting in an infra-red excess in the $L'$ band.  We also provide an additional astrometry point for 2014 June 12 and use the modelled companion mass combined with the RV and direct imaging data to place constraints on the orbital parameters for this companion.
\end{abstract}




\section{Introduction}
\label{sec:intro}

Until recently detections of white dwarf (WD) companions orbiting main sequence solar-type stars were relatively rare \citep{Holberg2013}, due to a combination of the high contrast between the WD and the primary and survey selection effects that favoured fainter, more distant FGK stars to avoid saturation.  However, a new analysis by \citet{Parsons2016} has recently identified nearly 900 bright main-sequence FGK stars with excess flux in the ultra-violet which is an indicator that the star is highly likely to have a WD companion.

These WD-FGK systems can provide a useful opportunity for detailed analysis of the characteristics of WDs. In particular, if they can be spatially resolved, then high contrast imaging and spectroscopy allows constraints to be placed on the temperatures and atmospheres of these objects, while calculations of the dynamical mass can be obtained from astrometric and radial velocity (RV) observations. These in turn can be used to test various theoretical WD models, such as cooling sequences and the  variation of the mass-radius relationship with WD temperature and composition \citep{Holberg2012}, both of which have important implications for the semi-empirical calculation of the initial-to-final mass ratio (IFMR) of WDs and their progenitors.

The HD 114174 (HIP 64150, GJ 9429) system consists of a WD orbiting a main sequence G star $26.14 \pm 0.37$ pc from the Sun \citep{vanLeeuwen2007}. The WD companion was discovered in 2013 \citep{Crepp2013} by the TRENDS high contrast imaging program, which uses RV measurements with long time baselines to identify stars that are likely to have wide orbit companions as promising targets for adaptive optics (AO) observations \citep{Crepp2012}.  The RV measurements of this system enabled a minimum mass of $0.26\pm0.01 M_{\odot}$ to be calculated which, combined with $J$ and $K_{s}$ photometry, indicated that the companion was a WD.  \citet{Crepp2013} classified the object as a hydrogen-rich WD with an effective temperature, $T_{\text{eff}} = 8200 \pm 4000$ K based on WD atmosphere model fits to their photometric data.  

\citet{Matthews2014} undertook follow-up observations of HD 114174 and obtained photometry of the companion in the $L'$ band and an upper limit from the non-detection of the companion in the $Y$ band.  Using the same modelling techniques and including the new photometry they found a much cooler effective temperature of $T_{\text{eff}} = 4260 \pm 360$ K.  This results in a cooling age for the WD of $7.77\pm0.24$~Gyr (with a corresponding total lifetime of $\sim$11 Gyr) which is significantly longer than the age of the primary, derived from isochronological and gyrochronological analyses as 4.7$^{+2.3}_{-2.6}$ Gyr and 4.0$^{+0.96}_{-1.09}$ Gyr respectively.  A more recent isochronal analysis by \citet{Tucci-Maia2016} results in a primary age of $6.406\pm0.656$ Gyr, however this is still too young to be consistent with the total age of the companion given the cooling age calculated by \citet{Matthews2014}.  Among the physical explanations suggested by \citet{Matthews2014} for this age discrepancy is the possibility that the host star has been spun up by slow, massive winds from the AGB progenitor of the WD making it appear younger \citep{Jeffries1996}. However, this does not account for the discrepancy with the isochronal age.  An alternative explanation suggested by \citet{Matthews2014} is that HD 114174 B has a faint undetected debris disk, although they consider this unlikely given its old age and cool temperature \citep{Kilic2009}. 

The system properties as listed on SIMBAD, the host star properties calculated by \citet{Crepp2013} and the TRENDS photometry for the companion are given in Table~\ref{tab:HD114174prop}. HD 114174 B has also been observed by GPI/Gemini in March 2014 as part of their astrometric calibration observations \citep{Konopacky2014} and by SPHERE/VLT as part of their commissioning phase in May 2014 \citep{Claudi2016}, in which the separation of the WD from the primary is given.  Neither project has published complete astrometry or photometry of this object yet.

\citet{Takeda2011} considered HD 114174 A as part of their study on Be depletion in solar-analogue stars and found that it was one of four stars, out of a sample of 118, that showed drastic Be depletion of more than $\sim$2 dex.  \citet{Desidera2016} consider the Be depletion of these stars in the context of detected or potential companions and they find that HD 114174 A has a mild super-solar abundance pattern for the $s$-process elements, indicating a likely interaction with the stellar winds from its companion during the companion's AGB phase.  They also suggest a link between a companion that has gone through the AGB phase and the significant Be depletion seen in these stars.

The layout of the paper is as follows; in $\S$\ref{sec:obs} we describe our observations of this system with Project 1640 (hereafter P1640), in $\S$\ref{sec:datred} we summarise our data reduction pipeline and the flux calibration of our final spectra and in $\S$\ref{sec:res} we present our astrometry analysis and spectral modelling. Finally, in $\S$\ref{sec:dis} we discuss our results and in  $\S$\ref{sec:sum} provide a summary of the paper.

\begin{table}
\centering
\small
\caption{HD 114174 properties}
\label{tab:HD114174prop}
\begin{threeparttable}
\begin{tabular}{lc}
\hline
\multicolumn{2}{c}{System$^a$}\\
\hline
Right ascension (J2000)      & 13 08 51.02          \\
Declination (J2000)          & +05 12 26.06         \\
$B$                          & 7.47                 \\
$V$                          & 6.80                 \\
$J$                          & 5.613 $\pm$ 0.026    \\
$H$                          & 5.312 $\pm$ 0.027    \\
$K_s$                        & 5.202 $\pm$ 0.023    \\
$d$ (pc)                     & 26.14 $\pm$ 0.37     \\
Proper motion (mas yr$^{-1}$)& 84.72 $\pm$ 0.59 E   \\
                             & -670.11 $\pm$ 0.47 N\\
\hline
\multicolumn{2}{c}{Host Star$^b$}\\
\hline
Mass ($M_{\odot}$)          & 1.05 $\pm$ 0.05     \\
Radius ($R_{\odot}$)        & 1.06                \\
Luminosity ($L_{\odot}$)    & 1.13                \\
Gyrochronological Age$^c$ (Gyr) & 4.0$^{+0.96}_{-1.09}$\\
Isochronal Age (Gyr)        & 4.7$^{+2.3}_{-2.6}$ \\
{[}Fe/H{]}                  & 0.07 $\pm$ 0.03     \\
log $g$ (cm s$^{-2}$)       & 4.51 $\pm$ 0.06     \\
$T_{\text{eff}}$ (K)      & 5781 $\pm$ 44       \\
Spectral Type               & G5 IV-V             \\
$v$ sin $i$ (km s$^{-1}$)   & 1.8 $\pm$ 0.5   \\
\hline
\multicolumn{2}{c}{Companion$^d$}\\
\hline
$Y$                  & $>$ 14.0 $\pm$ 0.7    \\
$J$                  & 16.06 $\pm$ 0.11      \\
$K_s$                & 15.94 $\pm$ 0.12      \\
$L'$                 & 15.30 $\pm$ 0.16      \\
\hline
\end{tabular}
\begin{tablenotes}
$^{a}$system parameters from SIMBAD; $^{b}$host star parameters from \citet{Crepp2013}; $^{c}$gyrochronological age from \citet{Matthews2014}; $^{d}$companion photometry from \citet{Crepp2013} and \citet{Matthews2014}.
\end{tablenotes}
\end{threeparttable}
\end{table}

\section{Observations}
\label{sec:obs}

The P1640 instrument is based at the 5.1 m Hale telescope at Palomar Observatory, equipped with the PALM-3000 (P3K) extreme adaptive optics system \citep{Dekany2013}.   The instrument, as described in \citet{Hinkley2011}, consists of an apodized pupil Lyot coronagraph, an internal wavefront calibration system \citep{Zhai2012} and an integral field spectrograph (IFS) with a 200 by 200 microlens array.  The near-IR spectra produced cover most of the $Y$, $J$ and $H$ bands over 32 wavelength channels.  The exact wavelength range and scale are recalculated for each individual observation as part of our wavelength calibration procedure.

We obtained observations of the HD 114174 system on three nights in June 2012 and on one night in June 2014.  For each set of observations we took a sequence of short (1.5 s) unocculted  `core' images, where the primary star is visible and unsaturated, which serve as our calibration images.  This was followed by a set of longer ($\sim$ 6-9 min) occulted exposures,where the light from the primary is blocked by placing it behind the occulting mask, allowing the fainter companion to be seen.  The dominant source of noise for coronagraphic observations is static or quasi-static speckle noise which cannot be significantly reduced through standards method such as increased exposure time and so has to be removed during post-processing \citep{Hinkley2007}.  For P1640 we find that sets of multiple, shorter exposures provide the best post-processing results. The exposure times, $t_{\text{exp}}$, and number of exposures, $N_{\text{exp}}$, depend on the target observed and on the observing conditions and are listed for HD 114174 in Table~\ref{tab:obs}. 

Seeing information is not available for the 2012-06 observations, however other metrics, for example the percentage of the total core counts enclosed within a given aperture and the fraction of stellar light blocked by the coronagraph in the occulted images, can be used as alternative indications of data quality. These metrics for our observations are given in Table~\ref{tab:obs}. The associated errors are calculated as the standard deviation across the exposures within each observing epoch and as such give an indication of the variability of the data across that epoch.

Binary standards were observed during the course of each observing run for astrometric calibration. For the 2014 observation astrometric spots for locating the primary star behind the occulting mask were added by applying a sinusoidal pattern to the AO system deformable mirror.   

\begin{figure}
 \centering
 \includegraphics[width = 85mm]{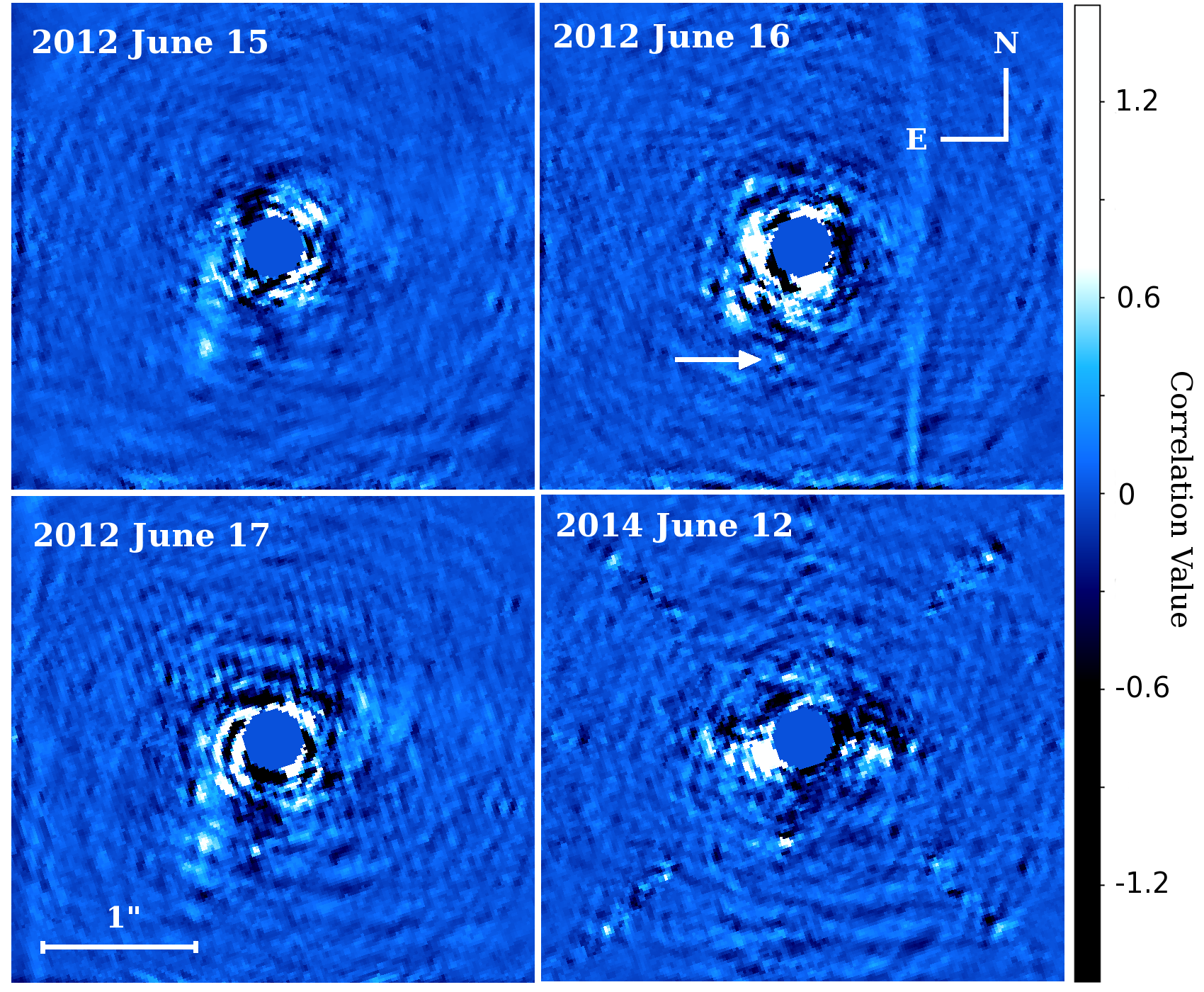}
 \caption{S4 detection maps for 2012 June 15, 2012 June 16, 2012 June 17 and 2014 June 12 observations of HD 114174. The arrow indicates the location of the HD 114174 B in the 2012 June 16 detection map and the companion can be seen in a similar location in the other maps.}
 \label{fig:HD114174B_maps}
\end{figure}

We see the companion in all our observations, as shown in Fig.~\ref{fig:HD114174B_maps}.  As the 2012 June 16 observations have the best data quality we focus on this data set for extracting a flux calibrated companion spectrum. However, the 2012 data sets lack grid spots, so we use the 2014 June 12 data set for our astrometry calculations.

\begin{table*}
	\centering
	\caption{Observations of HD 114174 with P1640}
	\label{tab:obs}
    \begin{threeparttable}
	\begin{tabular}{ccccccc}
		\hline
		Date & Julian Date & $N_{\text{exp}} \times t_{\text{exp}}$& Airmass & Seeing$^{a}$ & \% Flux in & \% Flux\\
        (UT) & (+2450000)& (s) && (arcsec) & Core Aperture & Suppression$^{b}$\\
		\hline
		2012 June 15 & 6093.7&5 $\times$ 366.6 & 1.158-1.234 &-         & 87.7 $\pm$ 5.6 & 72.5 $\pm$ 0.04\\
		2012 June 16 & 6094.7&3 $\times$ 549.9 & 1.155-1.269 &-          & 94.9 $\pm$ 2.4 & 78.9 $\pm$ 0.08\\
        2012 June 17 & 6095.7&2 $\times$ 366.6 & 1.140-1.190 &-         & 89.3 $\pm$ 7.0 & 44.8 $\pm$ 0.3\\
        2014 June 12 & 6820.7&9 $\times$ 557.7 & 1.136-1.258 &1.4-1.5   & 87.1 $\pm$ 5.5 & 70.9 $\pm$ 0.05\\
		\hline
	\end{tabular}
    \begin{tablenotes}
    $^{a}$The seeing listed is recorded by the Palomar 18-inch seeing monitor observing Polaris, and so does not precisely reflect the seeing conditions of our observations. $^{b}$Suppression of primary flux calculated as the total counts averaged across the core exposures minus the total counts averaged across occulted exposures and  expressed as a percentage of the average total core exposure counts.
    \end{tablenotes}
    \end{threeparttable}
\end{table*}

\section{Data Reduction}
\label{sec:datred}
Extensive data reduction procedures are required for combined coronagraphic/IFS observations.  These consist of three main steps; separating out the spatial and wavelength information from the initial 2D image into a 3D data cube, performing speckle suppression to disentangle the companion signal from the background speckle noise and calibration of the companion flux.

The P1640 data reduction procedure has been described in detail in previous papers  \citep{Zimmerman2011, Oppenheimer2013, Fergus2014}. However, a number of updates to our methodology have not yet been published, so we summarise the main aspects of the pipeline in the following section, before proceeding to describe a new calibration method for obtaining absolute flux measurements from our speckle suppressed data.

\subsection{Cube Extraction}
Each exposure results in a 2D `slope file' image consisting of 40,000 interleaved spectra (each corresponding to an individual microlens in the IFS) which are extracted into 3D data cubes measuring $250 \times 250 \times 32$ in $x \times y \times \lambda$, where $x$ and $y$ are the number of pixels in the detector spatial dimensions and $\lambda$ is the number of wavelength channels. A detailed description of the focal plane solution and spectral registration methods used in our cube extraction process  (\textit{P1640 Cube Extraction Pipeline}, PCXP) can be found in \citet{Zimmerman2011}.  However, our flux calculation method has been updated and is currently unpublished, so we provide a short summary here.

Our new cube extraction method aims to significantly reduce the effect of cross-talk by fitting the fluxes in three adjacent microlens spectra at the same time in an attempt to disentangle the flux contributions from each.  The flux is extracted for each individual wavelength channel at a time, corresponding to a single row of pixels within the spectrum of interest.  A block of 13$\times$5 pixels is shifted on a sub-pixel level to align the microlens spectra positions, known from the focal plane solution, with the pixel grid of the detector.  The flux in the central row of 13$\times$1 pixels, which covers the horizontal cross sections of three spectra is then fitted with three Gaussian profiles. The central Gaussian is matched with the known location of the spectrum being extracted and the location of the Gaussians on either side are determined from the focal plane solution.  Along with the locations, the widths of the Gaussians are pre-determined before modelling. There was found to be variation in the widths of the spectra ranging from approximately 0.75 to 1.1 pixels, due to the changing focus across the detector array. This sigma (FWHM) variation is accounted for using an extension of the focal plane modelling which measures the distribution of widths across a skyflat image and fits a smooth function to it, giving sigma as a function of microlens position.  With position and width fixed, the three spectra are modelled as function of their amplitudes.  The counts from the model of the central Gaussian are then integrated and assigned to the correct spatial pixel and wavelength in the data cube.  This is repeated for each wavelength in each spectrum in the focal plane.

\subsection{Cube Alignment}
The effects of dispersion within the cubes as well as the alignment between cubes is handled by the \textit{Cube Alignment Centering and Stacking} module (CACS; Nilsson et al., in prep.)  For data with grid spots the position of the primary behind the occulting mask is calculated as the intersection point of two lines connecting diagonally opposing grid spots. Each frame is shifted to place the primary at the centre of the image.  The individual cubes are then stacked to form a hypercube with dimensions $x \times y \times \lambda \times N_{\text{exp}}$.  

As our June 2012 data does not have grid spots the location of the star behind the occulting mask is not easy to determine.  We aligned each frame within a cube using theoretically calculated corrections based on the atmospheric dispersion equation from \citet{Roe2002}.  We then used the Radon Transformation method described in \citet{Pueyo2015} to calculate the location of the star behind the occulted mask for each individual cube. Using these calculated locations the individual cubes are then shifted and stacked to form the hypercube input for our speckle suppression code.

\subsection{Speckle Suppression}

Speckle suppression was performed using \textit{S4} \citep{Fergus2014}, which exploits the chromatic dependence of the speckles by using principle component analysis (PCA) to model them in a joint radius-wavelength space.  For each location in the image, given as a function of radius, $r$, and angle, $\theta$, an annulus of width $\delta r \sim 12$ pixels is divided into a test region of width $\delta\theta \sim 10$ pixels, located at ($r$,~$\theta$), and a training region consisting of the rest of the annulus.  The speckles in the training region are modelled as a linear combination of eigenvectors, which are then used to recreate the speckles in the test region. The speckles move radially outwards with wavelength and so have a primarily diagonal structure in the radius-wavelength space while the companion signal has a vertical structure.  The eigenvectors cannot easily reproduce this vertical companion signal and so it is left as a residual when the speckle model is subtracted from the data. The companion is easily visible in the final speckle suppressed images (Fig.~\ref{fig:HD114174B_maps}), which were produced by convolving the data with a white light model point spread function (PSF) to further reduce the speckle noise. 

\subsection{Spectral Extraction}
\label{ssec:spec}
The spectral extraction was done using the \textit{S4} spectral extraction code (\textit{S4s}).  This follows the same approach as used for the detection phase described above, however the modelling and subtraction is only performed for one given location, the region in which the companion has been detected, and a model companion spectrum is included and allowed to vary at the same time as the background speckles are fitted.  A summary of this process can be found in Appendix B2 of \citet{Oppenheimer2013} and will be described in more detail in Veicht at al. (in prep.).  The spectral extraction is highly dependent on the choice of input parameters, specifically, the patch size, $\Delta\theta$, and the number of principal components used, $N_{\text{PC}}$.  This extent of this dependence and methods of selecting these input parameters are discussed in Veicht at al. (in prep.).  The optimum parameters are contingent on the companion properties, such as brightness and location, as well as the data quality, so we briefly summarise the parameter selection for our 2012 June 16 observations of HD 114174 B below.

We used $\Delta\theta = 7$ which was found, when compared to lower values, to result in higher output S4 flux and smaller fractional errors, as well as improved spectral stability for varying $N_{\text{PC}}$ values when the companion is relatively bright compared to the background speckles. When selecting $N_{\text{PC}}$ we look for the first period of spectral stability when increasing $N_{\text{PC}}$ in the range 10-200 in steps of 10, (Veicht et al., in prep.).  This first stable region is the region where the speckles are well modelled and subtracted, but the number of principal components is not yet high enough to be able to partially model and subtract the companion flux.  The relative brightness of HD 114174 B means that the value of $N_{\text{PC}}$ is comparatively low and we found that the optimum $N_{\text{PC}}$ values were 50, 60, 70 for the 2012 June 16 data set.

\subsection{Spectral Calibration}
We used core images of HD 114174 A as calibrator images to correct for the effects of atmospheric and instrumental throughput on our final spectral shape.  We used the CACS code to correct the core images for instrumental and atmospheric dispersion, before using aperture photometry to obtain the raw spectrum of the calibrator. We used the spectral type from \citet{Crepp2013} and JHK photometry from \citet{Cutri2003} (as listed in SIMBAD and given in Table~\ref{tab:HD114174prop}) to select a template spectrum that matches the host star as closely as possible.  From a range of options from the Pickles \citep{Pickles1998} and IRTF catalogues \citep{Rayner2009}, we selected the Pickles 55 (G5IV) template spectrum.  The template spectrum was degraded to match the P1640 resolution and divided into the raw calibrator spectrum to give the spectral response function (SRF).  The companion spectrum output from S4 was then divided by this SRF to correct for the atmospheric transmission and instrument sensitivity.

\subsubsection{Error Calculation}
We used the laser and skyflat calibration observations taken at the start of each run to estimate the systematic errors introduced into our data by our pipeline,  specifically the cross-talk errors, which are wavelength dependent, and the pseudo-random spatial error on each spatial pixel due to uncertainties in the focal plane solution, image registration and modelling \citep{Zimmerman2011}.  This is combined with the shot noise, and the error on the template spectrum.  As we are using observations of HD 114174 A as the spectral calibrator, as opposed to a dedicated calibrator star with a well known NIR spectrum, we used the ratio between the IRTF G5V and Pickles G5IV template spectra, which are both close matches to the host star colours, to estimate the error that the choice of template spectrum introduces into the SRF \citep{Lewis2012}.  Finally, the errors on our S4 extracted companion spectrum were calculated using white light fake sources, inserted at the same radius as the companion, but at 50 different randomly selected $\theta$ values.  The error on the final spectrum is taken as the root mean square difference between the inserted and extracted spectra averaged across all locations and $N_{\text{PC}}$ values (Veicht et al., in prep.).

\subsubsection{Wavelength Calibration}
Neither PCXP nor S4 make any assumption about the absolute wavelength values of an observation.  However, for spectral calibration, in order to match the template spectrum to the calibration star aperture photometry spectrum and to the S4 extracted companion spectrum we need to assign our 32 wavelength channels, $f_i$, to the correct wavelength $\lambda_i$ via 
\begin{equation}
 \lambda_i = a(f_i -1) + b
\end{equation}
where $a$ is the scaling factor, or wavelength bin size, and $b$ is the offset.  

We calculate $a$ using our laser calibration observations, where 1310~nm and 1550~nm laser calibration slope files are co-added and extracted with PCXP. By equating the separation between the laser peaks in pixels to the known separation in wavelength we found $a$ equal to $27.59\pm0.47$ nm pix$^{-1}$ for the June 2012 epoch.  

To calculate the offsets for each exposure we use the P1640 filter transmission profile and a model NIR spectrum of the atmospheric transmission above Cerro Pachon (airmass = 1.0, water vapour column = 4.3mm) generated by the Gemini Observatory using the ATRAN atmospheric modelling software \citep{Lord1992}.  We combine these profiles with the stellar template spectrum and an estimation of the throughput of the instrument to create a model spectrum and then perform a weighted minimisation of the residuals between our model and our data to calculate the offsets for each individual core and occulted cube.  The atmospheric transmission profile is primarily dependant on the level of water vapour in the atmosphere.  As this is not recorded for our observations the atmospheric model we used was chosen to match our data as closely as possible, with a focus on the shape and depth of the water bands.

The offsets for each individual cube can vary by up to 10 nm due to variation in the spectral registration, but the wavelength range is approximately 940-1823~nm.  However, the low flux transmission due to the water band and the filter edges means that we exclude some of our data points in these regions as being unreliable. For the filter edges we exclude the three data points at either end of the spectrum, i.e. wavelength channels 1-3 and 30-32.  We also exclude the three data points centred around the main waterband feature at $\lambda\sim 1400$ nm, which correspond to wavelength channels 16-18.

\subsection{Flux Calibration}
\begin{figure}
 \centering
 \includegraphics[width = 80mm]{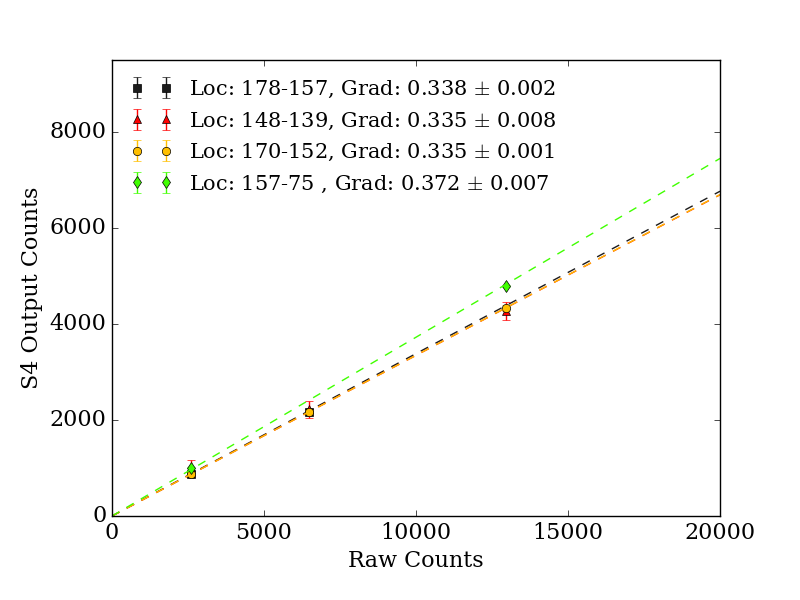}
 \caption{Flux calibration plots for 2012 June 16 data. The plotted points represent different fake source locations and brightnesses, with the brightness indicated by the Raw Count value on the $x$-axis and the location represented by the colour.  The gradients fitted to the different locations are included in the figure legend.}
 \label{fig:calibplot}
\end{figure}

The S4 output is given as relative flux in arbitrary units. In order to compare our spectrum to the previous photometric results we calculated a conversion factor  relating the output S4 counts to the raw counts in the original observations. Fake sources created using the primary PSF in the core cubes were inserted into the occulted cubes, at a range of locations and brightness, and then re-extracted using S4.  The linear relation between the brightness of an inserted companion and the output S4 spectrum flux for four different fake source locations in the 2012 June 16 data is shown in Figure \ref{fig:calibplot}.

Combining the gradients as a weighted mean, with the fitting errors taken as the weights, results in a conversion factor from S4 counts to raw counts of $0.336\pm0.016$ for the 2012 June 16 data set, where the error is the standard deviation across the gradients.

Having converted the companion spectrum into raw counts, the difference in magnitude between the primary and companion was calculated using,

\begin{align}
 \Delta J = -2.5\text{log}_{10}\bigg(\frac{\int f^{\text{B}}_{\lambda}S_J(\lambda) d\lambda}{\int f^{\text{A}}_{\lambda}S_J(\lambda) d\lambda} \bigg)
\end{align}

where $f^{\text{A}}_{\lambda}$ and $f^{\text{B}}_{\lambda}$ are the raw spectra for the host and companion respectively, which have been interpolated to match the wavelength grid of the filter profile, $S_J(\lambda)$.   Following \citet{Crepp2013} and \citet{Matthews2014} we used the MKO J band as the filter profile \citep{Tokunaga2002}.

The error contribution from the companion is calculated as a combination of the background shot noise (calculated from aperture photometry on the occulted cubes at the location of the companion), the S4 count to raw counts conversion errors and the average S4 extraction error across the J band wavelength range.  For the primary, the error is calculated as the shot noise plus the average of the PCXP and cross-talk error contributions over the J band range.  The total J band counts are calculated for each core cube individually and then combined using a weighted mean to give the final value used for calculating $\Delta J$.

For the 2012 June 16 data set we find $\Delta J = 10.33 \pm 0.24$, which is consistent with the previous result from \citet{Crepp2013} of $\Delta J = 10.48 \pm 0.11$. This results in a companion magnitude of $J_{MKO} = 15.91 \pm 0.24$, which can be converted into flux units using the the MKO Vega fluxes from from \citet{Tokunaga2005} resulting in $F_{\lambda} = 1.3001 \pm 0.2873 \times 10^{-22}$ W cm$^{-2}$ nm$^{-1}$ at the isophotal wavelength of 1250 nm.

In order to convert the companion spectrum into flux units, we corrected the spectrum using the SRF and then converted it to counts s$^{-1}$ nm$^{-1}$ by converting the S4 counts into raw counts and dividing by the wavelength scale of 27.59 nm per channel. We interpolated this to the filter wavelength, multiplied it by the filter profile and integrated to the obtain the total counts s$^{-1}$ in the J band. This was then equated to the total flux in the J band, obtained by multiplying $F_{\lambda}$ by the J band MKO bandwidth of 160 nm. The final P1640 spectrum in flux units compared to the TRENDS photometric points is shown in Fig~\ref{fig:HD114174bbfit}.

\section{Results}
\label{sec:res}
In this section we present modelling of our 2012 June 16 spectrum and astrometry for the 2014 June 12 data.  We did not calculate astrometry for our 2012 data as, even using the Radon transform method, the lack of grid spots results in very large errors on the position of the primary behind the occulting mask.  However, the TRENDS epochs already adequately cover this time period.

\subsection{Spectral Modelling}

We fitted hydrogen and helium atmosphere WD models to the 2012 June 16 flux calibrated spectrum, combined with the TRENDS $J$, $K$ and $L$ band photometry points. Only hydrogen and helium atmosphere models are considered as previous fits indicate that the WD is relatively cool and the low resolution of our spectrum precludes the possibility of fitting anything other than the overall slope. The hydrogen atmosphere models are described in \citet{Tremblay2011} and the Helium atmosphere models are described in \citet{Bergeron2011}. The model grids cover a range of effective temperatures from 1500 K to 40,000 K and a surface gravity range from $7.0 \leqslant \text{log}g \leqslant 9.0$.  Given the source distance, zero extinction is assumed.

The modelling is performed as follows. Initial guesses for $T_{\text{eff}}$ and $\text{log}g$ are provided for which the model grid is interpolated to produce the corresponding synthetic spectral energy distribution (SED). The normalisation factor between the SED model and the observed spectrum is then calculated, which is dependant only on the solid angle. Since the distance to the system is well known, the solid angle depends in turn only on the radius of the WD, $R_{\text{WD}}$.  This step is used to calculate an initial estimate for $T_{\text{eff}}$ and $R_{\text{WD}}$, from which chi-squared minimization is used to determine the best fit solutions for these values.  

The fitted radius is used to determine the mass and cooling age via WD evolutionary models based on those described in \citet{Fontaine2001} although using C/O cores, except for the lowest mass solutions, $M<0.4M_{\odot}$, where the appropriate He-core evolutionary models of \citet{Althaus2001} were incorporated. Generally the $\text{log}g$ value, calculated from the fitted mass and radius via $g = GM/R^2$, does not match the initial guess, so the fitting procedure is iterated until an internal consistency in $\text{log}g$ is achieved.

\begin{figure*}
 \centering
 \includegraphics[width = 180mm]{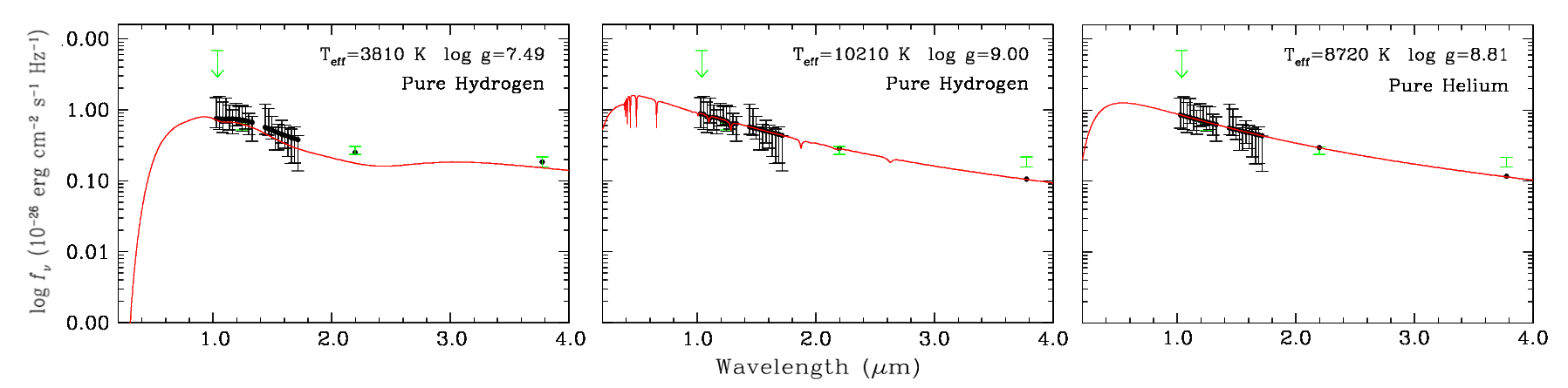}
 \caption{WD model fits to HD 114174 B spectrum.  The black and green error bars represent the P1640 and TRENDS measurements respectively, while the red line shows the model SED fit to the data. The black points represent the SED model at the P1640 and TRENDS wavelengths. }
 \label{fig:HD114174wdfit}
\end{figure*}

We ran the modelling for a range different initial effective temperatures and with the initial guess for the surface gravity as $\text{log}g = 8.0$.  The results for the combined P1640 spectrum and TRENDS photometry are given in Table~\ref{tab:WDmodels} and plotted in Fig.~\ref{fig:HD114174wdfit}.  The H fit was found to be dependent on the initial temperature, $T_{\text{init}}$. Fits with $T_{\text{init}} < 5000$~K converged to a very cool WD, where the effect of collisionally induced absorption allows the model to reconcile both the steepness of the P1640 spectrum and the TRENDS $K_s$ and $L'$ band photometry. For fits with $T_{\text{init}} > 5000$~K the model converges to a hot WD. This model fits the P1640 spectrum and TRENDS $K_s$ band well but cannot account for the flux seen in the $L'$ band photometry.  For all values of $T_{\text{init}}$ the He model resulted in a high temperature fit.

\begin{figure}
 \centering
 \includegraphics[width = 80mm]{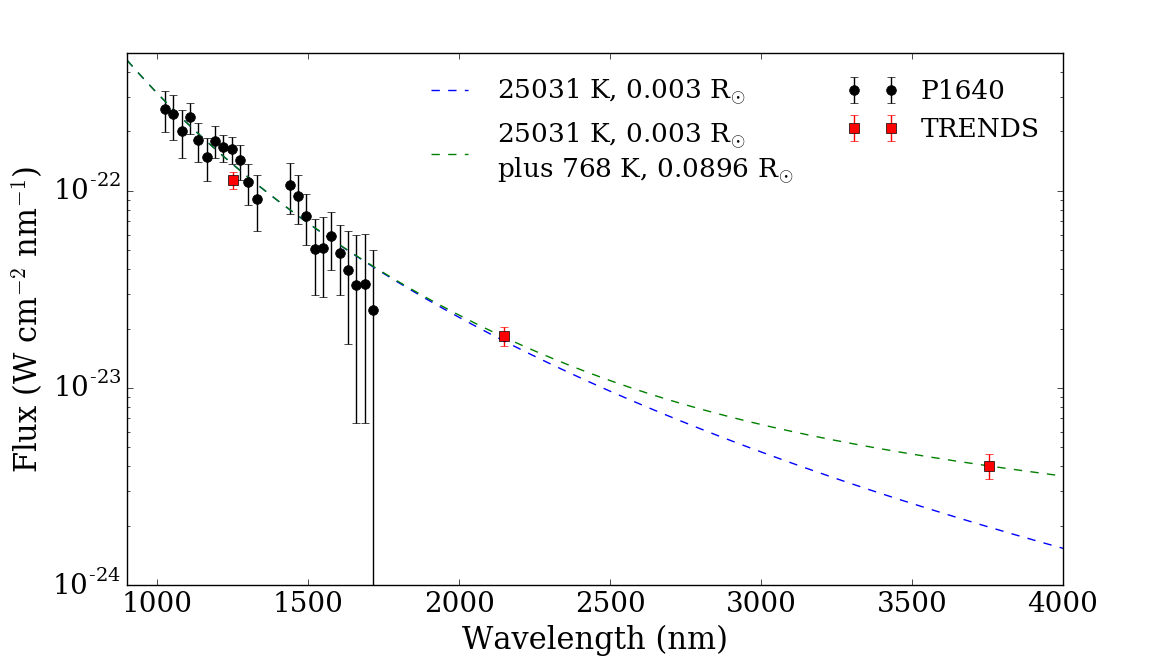}
 \caption{Blackbody fits to P1640 and TRENDS data, showing that the $L'$ band photometry can be modelled by a second cooler blackbody emission.}
 \label{fig:HD114174bbfit}
\end{figure}

The TRENDS $L'$ band flux is 1.6 and 1.8 times the flux for the He and high $T_{\text{init}}$ H models respectively, while the difference between the measured flux and the two models is 2.6 and 3.0 times the measurement error.  While it is possible that a systematic or measurement error has caused the high $L'$ band measurement, we consider it worthwhile to assume that the measurement represents a real IR excess compared to the He  and high $T_{\text{init}}$ H models and to consider possible causes. The most likely scenario is that the $L'$ excess is due to a second source.  To investigate this we fitted an initial blackbody to the P1640 spectrum, which resulted in a high temperature, although very badly constrained, fit of $25032\pm22757$ K.  We then used these fitted values in a two blackbody fit to the combined P1640 and TRENDS data, allowing the temperature and radius of the second blackbody to vary.  We found that the $L'$ band flux can be reproduced with a second blackbody with temperature $768\pm353$ K and radius 62346 $\pm$ 72899 km (Fig~\ref{fig:HD114174bbfit}).  

\begin{table*}
\centering
\caption{WD model fits for 2012-06-16 spectrum}
\label{tab:WDmodels}
\begin{tabular}{lccc}
\hline
Model			 & H ($T_{\text{init}} < 5000$)  & H ($T_{\text{init}} > 5000$)  & He ($T_{\text{init}} = $any)  \\
\hline
$T_{\text{eff}}$ (K)     & 3814 [3596,4032]    & 10212 [6720,13704]  & 8721 [6064,11377]   \\
Mass (M$_{\odot}$)       & 0.336 [0.325,0.346]  & 1.198 [1.192,1.204]  & 1.091 [1.083,1.100] \\
log$g$ (cm s$^{-2}$)                   & 7.49 [7.46,7.51]     & 9.00 [8.99,9.02]  & 8.81 [8.79,8.82]  \\
$R_{\text{WD}}$$\times 10^{-4}$ (R$_{\odot}$)  & 172.8 [170.4,175.3] & 57.1 [56.3,57.9]  & 68.2 [67.2,69.1]   \\
$M_{\text{bol}}$         & 15.36 [15.33,15.39]  & 13.49 [13.46,13.52]   & 13.79 [13.76,13.82]  \\
log($L$/$L_{\odot}$)     & -4.24 [-4.26,-4.23]  & -3.50 [-3.51,-3.48]   & -3.62 [-3.63,-3.60] \\
Age (Gyr)                &  9.170 [8.762,9.583] & 2.365 [2.364,2.363]   &  2.807 [2.796,2.815]\\
$\chi^2$                 & 12.88 & 19.44 & 17.85           \\  
\hline
\end{tabular}
\end{table*}

\subsection{Astrometry}

\begin{table}
\centering
\caption{Astrometry for HD 114174 B}
\small
\label{tab:TRENDSast}
\begin{threeparttable}
\begin{tabular}{llcc}
\hline
Project &Date         & $\rho$          & Position Angle   \\
&(UT)         & (mas)           & ($^{\circ}$)    \\
\hline
TRENDS& 2011 Feb 22$^a$ & $719.8 \pm 6.6$ & $171.8 \pm 0.5$  \\
TRENDS& 2012 Feb 2$^a$ & $701.1 \pm 5.0$ & $172.1 \pm 0.4$  \\
TRENDS& 2012 May 29$^a$ & $695.8 \pm 5.8$ & $170.5 \pm 0.6$   \\
TRENDS& 2012 Jun 4$^a$ & $692.1 \pm 8.7$ & $ 172.2 \pm 0.8$ \\
TRENDS& 2013 May 24$^b$ & $675 \pm 16 $  & $174.3 \pm 1.4$  \\ 
GPI& 2014 Mar 23$^c$ & $662.5 \pm 0.55 $  & \\ 
SPHERE& 2014-May$^d$  & 650 & \\ 
P1640& 2014 Jun 12$^e$ & $641\pm23$	& $172\pm0.6$\\
\hline
\end{tabular}
\begin{tablenotes}
\centering
$^{a}$\citet{Crepp2013}; $^{b}$\citet{Matthews2014}; $^{c}$ Calculated from the final plate scale and the observed angular separation in pixels listed in \citet{Konopacky2014}; $^{d}$\citet{Claudi2016}.$^{e}$Results from this paper.
\end{tablenotes}
\end{threeparttable}
\end{table}

\begin{figure*}
 \centering
 \includegraphics[width = 150mm]{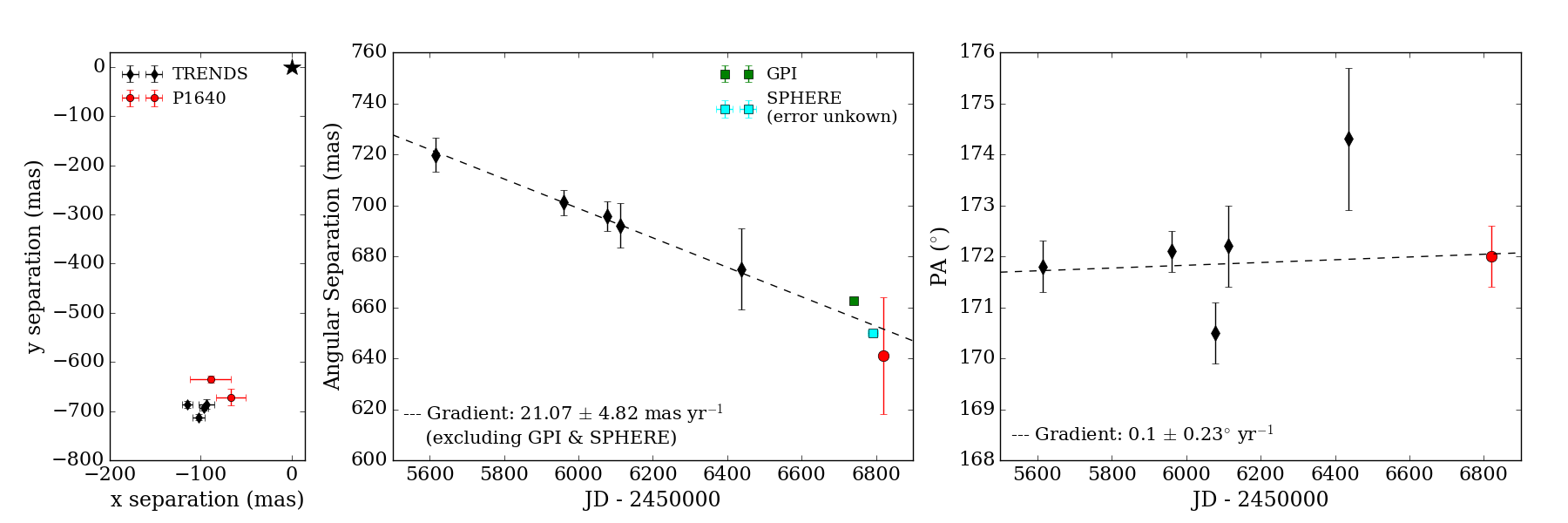}
 \caption{Astrometry fitting for HD 114174 B.  The black diamonds represent TRENDS astrometry measurements.  The red circle shows the P1640 2014-06-12 astrometry measurements and the blue and green squares show the GPI and SPHERE separations respectively. }
 \label{fig:HD114174B_locs}
\end{figure*}

During the 2014 June run we observed HIP 98001, HIP 88745 and HIP 107354 as calibration binaries, from which we calculated a plate scale of $18.90 \pm 0.61$mas pix$^{-1}$ and a position offset angle of $-71.68^{\circ} \pm 0.12^{\circ}$.

The position of the star behind the occulting mask is calculated as part of the CACS processing using the grid spots, with an error of 0.17 pixels.  The companion location is calculated by running \textit{S4s} in `planet shift' mode. This treats the companion location as a free parameter, modelling it at the same time as the spectrum and background speckles ($\S$\ref{ssec:spec}) and returning the calculated location as x-y pixel co-ordinates.  As the companion is not centred on a single pixel, \textit{S4s} is run using the four pixels with the highest correlation values as different starting locations.  The mean across these starting locations and across the three $N_\text{PC}$ values used to extract the spectrum is taken as the final x and y position.  The standard deviation is used as the uncertainty, resulting in errors of 0.47 and 0.27 pixels for the x and y coordinates respectively.

The separation and position angle calculated for HD 114174 B from the 2014 June 12 data is listed in Table~\ref{tab:TRENDSast} along with previously published values. From Fig.~\ref{fig:HD114174B_locs} it can be seen that no curvature is yet apparent, so we cannot yet constrain the orbital parameters via fitting either the directly imaged orbit or the RV curve.

However, we can combine the RV and direct imaging measurements with the companion's mass as calculated from the model fits to place restrictions on the possible orbital parameters. We run the following calculations using both the high and low mass fits from Table~\ref{tab:WDmodels} of 1.198 M$_{\odot}$ and 0.336 M$_{\odot}$ respectively. 

In order to calculate the orbital parameters, we first calculated the position, $\mathbf{r} = (x_B, y_B, z_B)$, and instantaneous velocity, $\mathbf{v} = (\dot{x}_B, \dot{y}_B, \dot{z}_B)$, vectors of the companion for a given epoch.  For our calculations we used the TRENDS separation and PA from 2012 June 4. The direct imaging measurements allow $x_B, y_B, \dot{x}_B$ and $\dot{y}_B$ to be calculated via, 

\begin{align}
 x &= R~\text{cos}\phi \\
 y &= R~\text{sin}\phi \\
 \dot{x} &= \dot{R}~\text{cos}\phi - y\dot{\phi} \\
 \dot{y} &= \dot{R}~\text{sin}\phi + x\dot{\phi}  
\end{align}

where $R =\frac{\rho}{\pi_A}$ is the projected separation in AU, with $\rho$ being the measured separation and $\pi_A$ being the parallax of HD 11414 A . $\dot{R}$ and $\dot{\phi}$ are the change in separation and PA.  As we do not see any curvature in the orbit we calculated these values using a straight line fit to the combined P1640 and TRENDS astrometry (Fig.~\ref{fig:HD114174B_locs}), resulting in $\dot{R} = 21.07\pm4.82$~mas~yr$^{-1}$ and $\dot{\phi} = 0.1\pm0.23^{\circ}$~yr$^{-1}$.

We used the calculations given in \citet{Howard2010} to calculate the physical separation, $r$, between the secondary and the primary using the fitted RV trend of $61.1\pm0.1$~m~s$^{-1}$~yr$^{-1}$ from \citet{Crepp2013} as the instantaneous Doppler acceleration, $\ddot{z_A}$.  Defining $\vartheta$ as the angle between the line connecting the primary to the secondary and the light of sight to the primary, such that $\vartheta = 0$ corresponds to the point at which the secondary is directly behind the primary, then

\begin{align}
\ddot{z_A} = \frac{Gm_B}{r^2}\text{cos}\vartheta,
\label{eq:zb1}
\end{align}
where $G$ is the gravitational constant and $m_B$ is the mass of the secondary, and
\begin{align}
r \text{sin}\vartheta = \frac{\rho}{\pi_A} = R.
\label{eq:zb2}
\end{align}

From this we calculated the line of sight position of the companion as 
\begin{align}
z_B = \pm \sqrt{r^2 - R^2}
\label{eq:zb3}
\end{align}
with a degeneracy as to whether the companion is in front of or behind the primary.  We also used the RV measurements to calculate $\dot{z}_B$ using,
\begin{equation}
 \dot{z_B} = -\frac{\dot{z}_A m_A}{m_B}
\end{equation}
where $m_A$ is the mass of the primary and $\dot{z_A}$ is the instantaneous acceleration of the primary, calculated by extrapolating the RV measurements given in \citet{Crepp2013} to the chosen direct imaging epoch.

A Keplerian orbit can be completely described by five parameters:semi-major axis, $a$, eccentricity, $e$, inclination, $i$, longitude of ascending node, $\Omega$ and argument of pericentre, $\omega$, with the position of the body along its orbit described by the true anomaly, $f$. From the position and velocity of the secondary\footnote{In the following calculations the position and velocity vectors refer only to the companion so for clarity the subscripts have been omitted.} and defining $\mathbf{h} =  \mathbf{r} \times \mathbf{v} = (h_X, h_Y, h_Z)$ we calculated these orbital parameters using the equations given in \citet[Appendix B]{Pearce2015},

\begingroup
\addtolength{\jot}{1em}
\begin{align}
 &a = \Big(\frac{2}{r} - \frac{v^2}{\mu}\Big)^{-1} \\
 &e = \sqrt{1=\frac{h^2}{a\mu}},~~\text{where}~\mu = G(m_A + m_B) \\
 &\text{cos}~i = \Big(\frac{h_Z}{h}\Big) \\
 &\Omega = \text{atan2}~(h_X,-h_Y) 
\end{align}
\endgroup
\begin{align}
 &f = \text{atan2}~\Big(\frac{p\dot{r}}{he}, \frac{p-r}{re}\Big), \\
 &~~~~~~~~\text{where}~~ p = a(1-e^2) ~~\text{and}~~ \dot{r} = \text{sng}(\mathbf{r}\cdot\mathbf{r})~\sqrt{v^2-\frac{h^2}{r^2}} \nonumber
\end{align}
\begin{align}
 &w =  \theta - f = \text{atan2}~(\text{sin}\theta,\text{cos}\theta)  -f, \\
 &~~~~~~~~\text{where}~~ \text{sin}\theta = \frac{z}{\text{sin}i} ~~\text{and}~~ 
 \text{cos}\theta = \text{sec}\Omega~\Big(\frac{x}{r} +\frac{z~\text{sin}\Omega}{r~\text{tan}i} \Big) \nonumber
\end{align}

\begin{figure}
 \centering
 \includegraphics[width = 85mm]{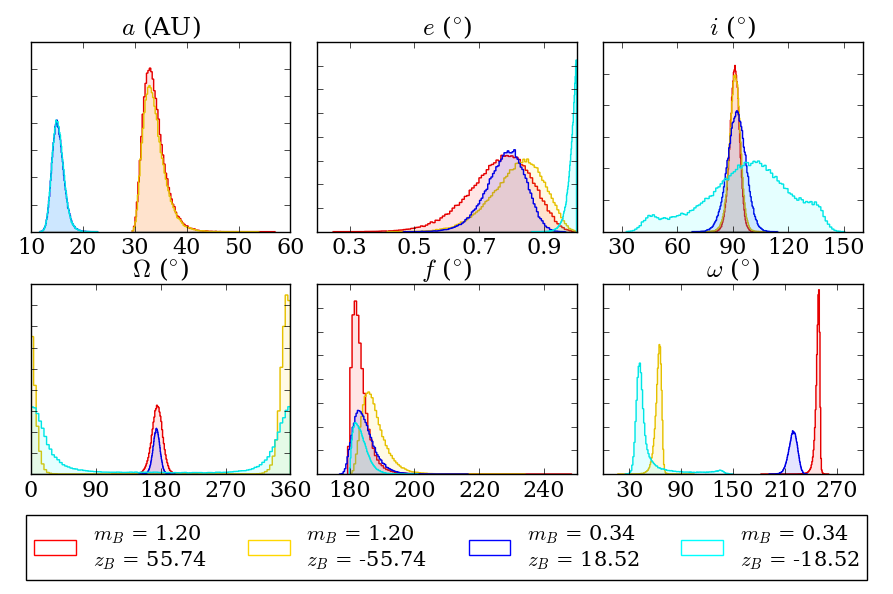}
 \caption{Orbital parameter distributions for high and low mass, positive and negative $z_B$ scenarios. The values for $z_B$ are calculated from Eqs. \ref{eq:zb1}, \ref{eq:zb2} and \ref{eq:zb3} and as such are dependant upon the value of $m_B$.}
 \label{fig:HD114174B_orb}
\end{figure}

Table~\ref{tab:orbparams} gives the median and 68\% confidence limits from these simulations for the two different WD masses and for positive and negative $z_B$, calculated using a Monte Carlo simulation drawing the $\mathbf{r}$ and $\mathbf{v}$ components from a normal distribution with the standard deviation equal to the calculated errors on each component.  The distributions are also shown in Fig.~\ref{fig:HD114174B_orb}, illustrating the dependency of the calculated parameters on the choice of WD mass and the sign of $z_B$.

\begin{table*}
\centering
\caption{Orbital parameters for high and low mass WD models with positive and negative $z_B$.}
\label{tab:orbparams}
\begin{tabular}{lcccc}
\hline
Mass ($M_{\odot}$)    & \multicolumn{2}{c}{1.198 $\pm$ 0.006}                     & \multicolumn{2}{c}{0.336 $\pm$ 0.014}                     \\
$z_B$ (AU)            & + 55.74                     & - 55.74                     & + 18.52                     & - 18.52                     \\
\hline
$a$ (AU)              & 33.44 {[}31.80, 35.74{]}    & 33.44 {[}31.81, 35.75{]}    & 15.16 {[}14.12, 16.38{]}    & 15.17 {[}14.13, 16.38{]}    \\
$e$                   & 0.77 {[}0.67, 0.86{]}       & 0.82 {[}0.74, 0.90{]}       & 0.79 {[}0.72, 0.84{]}       & $> $0.981                   \\
$i$ ($^{\circ}$)      & 91.02 {[}88.39, 93.53{]}    & 91.18 {[}88.18, 94.02{]}    & 92.00 {[}86.96, 96.88{]}    & 98.86 {[}74.39, 120.60{]}   \\
$\Omega$ ($^{\circ}$) & 175.15 {[}167.57, 182.45{]} & 355.61 {[}346.97,363.86{]}  & 174.12 {[}169.33, 178.80{]} & 361.42 {[}336.22, 388.92{]} \\
$f$ ($^{\circ}$)      & 182.72 {[}181.05, 185.55{]} & 186.94 {[}181.17, 191.12{]} & 184.00 {[}181.34, 187.71{]} & 183.11 {[}181.37, 185.66{]} \\
$\omega$ ($^{\circ}$) & 248.41 {[}245.52, 250.17{]} & 64.23 {[}60.03, 67.14{]}    & 219.71 {[}214.47, 224.43{]} & 44.16 {[}39.34, 60.31{]}  \\ 
\hline
\end{tabular}
\end{table*}

The calculated parameters for the $m_B = 0.34$~M$_{\odot}$ negative $z_B$ scenario stand out, as $i$ is not well constrained and the most likely eccentricity is 1. However, for the remaining scenarios a bound orbit with $e > 0.5$, an almost edge-on inclination and a true anomaly between 180$^{\circ}$ and 190$^{\circ}$ are all common features.  The fitted values of $\Omega$ and $\omega$ are primarily dependent on the sign of $z_B$, with $\Omega\sim180^{\circ}$, $\omega\sim220$-$250^{\circ}$ for positive $z_B$ and $\Omega\sim360^{\circ}$, $\omega\sim40$-$70^{\circ}$ for negative $z_B$.  The semi-major axis is the parameter most dependent on mass, ranging from $15.2^{+1.2}_{-1.1}$~AU for $m_B = 0.34$~M$_{\odot}$ to $33.4^{+2.3}_{-1.6}$~AU for $m_B = 1.20$~M$_{\odot}$, with corresponding orbital periods of $50.2^{+6.1}_{-5.1}$ years and $129.1^{+13.5}_{-9.5}$ years respectively.

\section{Discussion}
\label{sec:dis}
While our 2012 June 16 spectrum is consistent with the TRENDS $J$ band photometry, the extension of the spectrum into the $Y$ and $H$ bands allows further constraints to be placed on the characteristics of the companion. If we assume that the $L'$ measurement is correct then the steepness of the P1640 spectrum has restricted the modelling options to two possible scenarios.  The object could be a very cool WD with a H atmosphere, where the temperature is low enough ($\sim 4000$ K) that the effect of collisionally induced absorption can be seen in the spectral shape.  

This scenario allows the WD fitted spectrum to be consistent with the TRENDS $L$ band photometry and is similar to the results from \citet{Matthews2014}.  However, it pushes the fitted mass down to below $0.4 M_{\odot}$. This requires us to consider the possibility of a He-core WD, as standard evolutionary theory states that for single low- or intermediate-mass stars following the standard evolutionary tracks the minimum mass of the hydrogen core required for helium ignition and the production of a CO-core WD is around $0.5 M_{\odot}$ \citep{Prada-Moroni2009}.  As the cooling time is inversely proportional to the atomic weight of the interior ions this results in a much longer cooling age of 9.17 Gyr, further increasing the age discrepancy between the companion and the primary.  This long cooling age also indicates that a very high progenitor mass is required if the total lifetime of the system is not to exceed the age of the Galactic disk ($\sim$10~Gyr).  

Such a low WD mass also requires the progenitor to have undergone significant mass loss during the RGB phase, most likely via common envelope (CE) evolution with its companion (now the primary in the system) \citep{Althaus2001}.  Although HD 114174 A does show some evidence of interactions with the WD progenitor, the usual CE phase formation would result in a close binary with an orbital period on the order of days \citep{Brown2011}. However a number of single low-mass He-core WDs have been discovered (although none so far below 0.39~M$_{\odot}$), which cannot have evolved via a close binary system, and their proposed formation scenarios may be applicable to this system. For example, \citet{Nelemans1998} suggest that capture of a planet or brown dwarf orbiting the progenitor during the AGB phase can expel the envelope of the giant star, resulting in a low mass He-core WD.  Increasing detections of planets in binary systems, with stellar separations down to $\sim$20 AU \citep{Desidera2007, Mugrauer2009}, indicate that this evolutionary method is not impossible. 

It has been demonstrated that very specific evolutionary scenarios involving mass loss in a close binary could result in a WD with a mass as low as $0.33 M_{\odot}$ with a CO-core rather than a He core \citep{Prada-Moroni2009}. Low temperature CO-core fits to our spectrum result in a cooling age of $\sim$ 5 Gyr, which is more reconcilable with the age of the primary.  However, our calculated orbital parameters for this system, in particular the high eccentricity, make evolution via a close binary (which would probably have circularised the orbit) unlikely without some interaction with a third body.

Alternatively, the P1640 spectrum is fit relatively well by a higher temperature H or He WD of  $\sim 9500$ K, similar to that initially reported by \citet{Crepp2013}.  However, in order to remain consistent with the $L'$ band photometry this scenario requires a second much cooler body, such as a debris disk or brown dwarf (BD), as the source of the IR excess.  Both BDs and debris disks have been detected around WDs \citep[e.g.][]{Steele2013, Bergfors2014}, indicating that planetary systems and other companions can survive the evolution of a WD progenitor during its RGB phase.  

The age of the HD 114174 system suggests that a debris disk is unlikely.  The fraction of observable disks is correlated with the temperature of the WD (as an indicator of its age), and drops sharply at around $T_{\text{eff}}\approx 10000$~K, with only two WDs with $T_{\text{eff}}\lesssim 10000$~K and ages of $\approx 1$~Gyr so far discovered to have weak IR excesses indicative of narrow dust rings \citep{Bergfors2014}.  Considering the possibility of a BD companion, although there are only a handful of close WD-BD binaries \citep{Casewell2014} and none with cooling ages longer that $\sim1$~Gyr, it is not impossible that such a system could persist for longer in a stable configuration. Using Eq.~1 from \citet{Holman1999} and the orbital parameters for the high mass scenario given in Table~\ref{tab:orbparams} we estimate the critical radius to be $\sim1.2$~AU, outside of which stable orbits cannot exist.

This higher temperature fit also constrains the radius to be very small ($\sim 0.0057$~R$_{\odot} - 0.0068$~R$_{\odot}$) and the mass to be correspondingly high at around 1.09-1.2 M$_{\odot}$.  This would place HD 114174 B at the extremely high mass end of the known WD population.  For comparison, Sirius B is one of the highest mass known WDs orbiting a MS star, with a mass and radius of approximately 1.02 M$_{\odot}$ and 0.0081R$_{\odot}$ respectively \citep{Barstow2005}. Additionally, the Sloan Digital Sky Survey and Palomar Green Survey place the number of field WDs with masses greater than 1.05 M$_{\odot}$ at 1.5\% and 2.6\% respectively \citep{Cummings2016}. 

If the $L'$ excess is not real, this higher temperature fit is the most likely scenario.  Although it no longer requires the addition of a debris disk or BD, the unusually low radius and high mass still make the companion interesting and could provide a useful high mass data point for IFMR calculations.

\section{Summary} \label{sec:sum}
We have presented our data reduction and calibration of P1640 observations of the HD 114174 system.  Our 2014 June 12 observations extend the astrometry available for this system.  We place limits on the orbital parameters and show that the inclination is almost exactly edge-on and that the eccentricity is greater than 0.5 regardless of which WD model is assumed.

Our infra-red spectrum of HD 114174 B, when combined with previous TRENDS photometry in the $J$, $K_s$ and $L'$ bands indicates that two broad scenarios are possible to explain the current data.  Either HD 114174 B is an extremely cool, low mass, H atmosphere WD with $T_{\text{eff}}\sim3800$ K, requiring a very specific evolutionary path to account for the low mass, and, if assuming a He-core, the age discrepancy with the primary.  Alternatively, HD 114174 B could be an unusually high mass, hot WD probably with a debris disk or brown dwarf companion.  Observations in either the visible or UV would be most useful in order to differentiate between the different models, ideally with a further observation in the $L'$ band to confirm the previous photometry. Additionally, the mass of the companion can be constrained once suitable curvature is seen in the RV observations.

\section*{Acknowledgements}
We thank Tim Pearce, Grant Kennedy and Matthew Kenworthy for their helpful discussions concerning the astrometry of HD 114174 B.  We also thank Simon Hodgkin for his input concerning the data reduction and calibration of our data for this object and the referee for their detailed and helpful comments.

Project 1640 is funded by National Science Foundation grants AST-0520822, AST-0804417, and AST-0908484. In addition, part of this work was performed under a contract with the California Institute of Technology funded by NASA through the Sagan Fellowship Program and part of the research in this paper was carried out at the Jet Propulsion Laboratory, California Institute of Technology, under a
contract with the National Aeronautics and Space Administration (NASA). 

EB is supported by an STFC studentship, JA is supported by the Laboratory for Physical Sciences, College Park, and MD, through the National Physical Science Consortium graduate fellowship program. JC was supported by the U.S. National Science Foundation under Award No. 1009203. RN was funded by the Swedish Research Council's International Postdoctoral Grant No. 637-2013-474. JRC acknowledges support from NASA Origins Grant NNX13AB03G and the NASA Early Career program. The members of the Project 1640 team are also grateful for support from the Cordelia Corporation, Hilary and Ethel Lipsitz, the Vincent AstorFund, Judy Vale, Andrew Goodwin, and an anonymous donor.

This paper is based on observations obtained at the Hale Telescope, Palomar Observatory.  This research made use of the Washington Double Star Catalogue maintained at the U.S. Naval Observatory, the SIMBAD database, operated by the CDS in Strasbourg, France, NASA's Astrophysics Data System and data from ESA's Gaia mission. 

Facilities: Hale (Project 1640)




\bibliographystyle{mnras}
\bibliography{thesis} 






\bsp	
\label{lastpage}
\end{document}